\documentclass[twocolumn,showpacs,preprintnumbers,amsmath,amssymb,aps,superscriptaddress]{revtex4-1}
\usepackage{graphicx}
\usepackage{epstopdf}           
\usepackage{dcolumn}
\usepackage{bm}
\usepackage[utf8]{inputenc} \usepackage[T1]{fontenc} % special letters
\usepackage{nicefrac}
\usepackage{xfrac}
\usepackage{lmodern}
\usepackage{float}
\usepackage{xcolor}
\usepackage{siunitx}

\usepackage{hyperref}

\usepackage{braket}

\begin{document}

\title{Spin beats in the photoluminescence polarization dynamics of charged excitons in InP/(In,Ga)P quantum dots in presence of nuclear quadrupole interaction}

	\author{S.~V.~Nekrasov}
	\email{nekrasov108@yandex.ru}
	\affiliation{Ioffe Institute, Russian Academy of Sciences, 194021 St.~Petersburg, Russia}
	
	\author{I.~A.~Akimov}
	\affiliation{Ioffe Institute, Russian Academy of Sciences, 194021 St.~Petersburg, Russia}
	\affiliation{Experimentelle Physik 2, Technische Universit\"at Dortmund, 44221 Dortmund, Germany}
	
	\author{Yu.~G.~Kusrayev}
	\affiliation{Ioffe Institute, Russian Academy of Sciences, 194021 St.~Petersburg, Russia}
	
	\author{D.~R.~Yakovlev}
	\affiliation{Ioffe Institute, Russian Academy of Sciences, 194021 St.~Petersburg, Russia}
	\affiliation{Experimentelle Physik 2, Technische Universit\"at Dortmund, 44221 Dortmund, Germany}
	
	\author{M.~Bayer}
	\affiliation{Ioffe Institute, Russian Academy of Sciences, 194021 St.~Petersburg, Russia}
	\affiliation{Experimentelle Physik 2, Technische Universit\"at Dortmund, 44221 Dortmund, Germany}

\begin{abstract}
The spin dynamics of positively (X$^{+}$) and negatively (X$^{-}$) charged excitons in InP/In$_{0.48}$Ga$_{0.52}$P quantum dots subject to a magnetic field is studied. We find that a characteristic feature of the system under study is the presence of nuclear quadrupole interaction, which leads to stabilization of the nuclear and electron spins in a quantum dot in zero external magnetic field. In detail, the nuclear quadrupole interaction leads to pinning of the Overhauser field along the quadrupole axis, which is close to the growth axis of the heterostructure. The nuclear effects are observed only when resident electrons are confined in the quantum dots, i.e. for X$^{-}$ trion photoexcitation. The presence of X$^{-}$ and X$^{+}$ trion contributions to the photoluminescence together with the quadrupole interaction significantly affects the dynamics of optical orientation in Voigt magnetic field. In absence of dynamic nuclear spin polarization the time evolution of the photoluminescence polarization was fitted by a form which describes the electron spin relaxation in ``frozen'' nuclear field fluctuations. In relatively large external magnetic fields exceeding 60 mT good agreement between theory and experiment is achieved.

\end{abstract}

\maketitle
\section{Introduction}
\label{sec:intro}

Quantum dots (QDs) are promising objects for use in spintronics devices~\cite{Dyakonov, Bracker} because the motion-related mechanisms of carrier spin relaxation that are important in bulk~\cite{OO3} are suppressed in QDs~\cite{Khaet}. The practical requirement of long-lived carrier spin coherence calls for studying effects that stabilize spins in nanostructures.  Here we investigate one such effect, the nuclear quadrupole interaction~\cite{Quadr}.

A nucleus with a spin exceeding 1/2 has a nonzero electric quadrupole moment~\cite{Slichter}. The presence of an electric field gradient across the nucleus location leads to the nuclear quadrupole interaction. For example, in bulk (Al,Ga)As semiconductors an electric field gradient appears when the gallium ions are replaced by aluminum ions~\cite{OO5}. In case of InP/(In,Ga)P QDs the field gradient is caused by a significant lattice deformation which is generated by the large mismatch of InP and (In,Ga)P lattice constants, reported to be 3.7\% between InP and In$_{0.485}$Ga$_{0.515}$P in Ref.~\cite{Kurt}. The deformation occurs at the heterointerface, while the deformation axis is aligned with the QDs growth axis (in our case the [001] axis). The quadrupole interaction affects the spin state of the nucleus, but the projection of the nuclear spin on the main axis of the quadrupole interaction [001] is preserved. If the strength of the electron and nuclear spins hyperfine interaction is significant, the nuclear quadrupole interaction will also affect the electron spin. This effect can be probed by means of polarized photoluminescence (PL) as used in the present study. 

The dynamic nuclear polarization (DNP) in presence of strong nuclear quadrupole interaction was previously studied in bulk (Al,Ga)As~\cite{OO5}. Also, the effect of the quadrupole interaction on the spin systems of electrons and nuclei in InP/(In,Ga)P QDs was observed~\cite{Quadr, Cheh}. Steady-state magnetic field studies of the negatively charged exciton (X$^{-}$ trion) PL polarization were carried out in Ref.~\cite{Quadr}, and the dynamics of nuclear spin polarization under circular polarized excitation in singly charged and neutral quantum dots in Faraday magnetic field were studied in Ref.~\cite{Cheh}. 

It should be noted that unlike for positively charged excitons (X$^{+}$ trions)~\cite{XPlus} and neutral excitons~\cite{X0}, the circular polarization dynamics of the negatively charged exciton PL does not exhibit oscillations in Voigt magnetic field, reported for InP/(In,Ga)P QDs in Ref.~\cite{X0, My}. The absence of these oscillations is discussed in Section~\ref{subsec:Num}. In case of the neutral excitons the anisotropic exchange interaction of the electron and hole spins~\cite{AEI} becomes relevant in presence of an anisotropy in the QD plane. As a result, the spin relaxation time of the carriers is significantly reduced~\cite{FastRel}. Thus, studying the optical orientation dynamics in Voigt magnetic field is most informative for the X$^{+}$ trion PL.

In this paper we study the effect of nuclear quadrupole interaction on the spin dynamics of an ensemble of InP/(In,Ga)P QDs that is fractionally positively charged (X$^{+}$), negatively charged (X$^{-}$) and charge neutral. The PL circular polarization dynamics and the steady-state PL polarization subject to magnetic field applied either in Voigt or in Faraday geometry are investigated. In the optical orientation dynamics in Voigt magnetic field pronounced spin beats are observed, which we study in presence of DNP and nuclear quadrupole interaction. In particular, we find that the PL circular polarization oscillates not around zero value, as typically observed~\cite{ZeroOsc1, ZeroOsc2}, but around a polarization contribution that monotonically decays with time. We show that this behavior can be explained by the simultaneous contributions of X$^{-}$ and X$^{+}$ trions to the PL. 

In order to quantitatively describe the PL polarization dynamics, it is necessary to determine whether the spin relaxation takes place in the limit of long~\cite{Merc} or short~\cite{OO3} correlation time: It is known that any mechanism of spin relaxation can be considered in terms of effective magnetic field fluctuations acting on the spin. An important characteristic of these fields is the correlation time ($\tau_c$), the period in time during which the field fluctuations remain unchanged. There are two extreme limits of spin relaxation. The limit of long correlation time (``frozen'' fields) corresponds to the condition $\Omega_f \tau_c \gg 1$, where $\Omega_f$ is a spin precession frequency in a field fluctuation. The condition $\Omega_f \tau_c \ll 1$ corresponds to the limit of short correlation time. We show that in the studied QDs the electron spins relax in  nuclear field fluctuations with a long correlation time. The time dependencies of the PL polarization (in absence of DNP) for different magnetic field strengths were described within the approach developed in Ref.~\cite{Merc}. 

\section{Experimental details}
\label{sec:ExpDetails}

The single layer of lens-shaped, self-organized InP quantum dots embedded in a In$_{0.48}$Ga$_{0.52}$P matrix was grown by metalorganic vapour-phase epitaxy on a (001) GaAs substrate. The QDs have a bimodal size distribution: The dots in one group have average sizes of about $100 \times 5$ nm$^{2}$ (diameter $\times$ height) and those in the second group have sizes of  $133 \times 20$ nm$^{2}$. The QDs were covered with a 40~nm In$_{0.48}$Ga$_{0.52}$P cap layer. No wetting layer is formed in these samples. A detailed description of the structure is given in~\cite{Samp}, denoted there as sample (i).

Studies of the PL intensity and polarization were carried out both in continuous wave mode (CW) and pulsed regime (PR) with time resolution. The sample was placed in a cryostat with liquid helium at 2~K temperature (CW) or with helium vapor at 6~K temperature (PR). The external magnetic field, $B$, was generated by an electromagnet ($B = 0-250$~mT) (CW) or superconducting coils ($B = 0-400$~mT) (PR). The PL was excited by Ti:Sph lasers (1.77 eV central photon energy) with a power density of about 75 W/cm$^2$, operated in continuous wave or pulsed mode. In the latter case, optical pulses with a duration of 150 fs were generated by a self-mode-locked oscillator at a repetition frequency of 75 MHz. The laser light was circularly polarized, and its direction approximately coincided with the sample growth axis. The PL was collected in ``reflection'' geometry, and the degree of its circular polarization, which is defined by
\begin{align}
\rho_c=\frac{I^{+} - I^{-}}{I^{+} + I^{-}},
\end{align}
was measured, where $I^{+}/I^{-}$ are the intensities of the PL components which polarization coincides/is opposite to the exciting light polarization. We note that when PL is excited with circular polarized light, dynamic polarization of nuclear spins may occur through the hyperfine interaction with the spin polarized electrons. When it was required to exclude DNP, a photoelastic modulator (CW) was placed in the excitation path modulating the polarization of light between $\sigma^{+}$ and $\sigma^{-}$ at a frequency of 26.61~kHz. In the PR studies an electro-optic modulator (EOM) was used (16~kHz). Due to the modulation the fast changes in the direction of the electron spin orientation prevent the build-up of DNP~\cite{OO5}. Finally, after passing through a double- (CW) or single-grating (PR) monochromator the PL was detected with an avalanche photodiode (CW) or with a streak camera (PR). In the latter case, the setup time resolution was about 30~ps.

In the CW regime, the PL intensities were measured using a two channel photon counting unit. In the PR regime, the required $\sigma^+$ or $\sigma^-$ polarization was selected manually in the detection path and the corresponding transients $I^+(t)$ or $I^-(t)$ were accumulated. Here, the record time range corresponded to a window of about 2~ns, and accumulation is synchronized with the repetition frequency of the laser (75~MHz). When the excitation polarization was modulated, the EOM was synchronized with the streak camera using a blanking unit. The streak camera blanking unit allowed us to provide an additional time filtering of the accumulated transients on a slow $\SI{}{\micro\s}$ timescale. Here, the PL time dependencies in a 2~ns window were measured only for less than a half of the EOM period (less than $\SI{20}{\micro\s}$), when the excitation polarization was constant ($\sigma^+$). Therefore, the PL polarization degree was analyzed only during the specified time interval ($\sigma^+$ excitation).

\section{Experimental results}
\label{sec:ER}

\begin{figure}[!h]
\centering
\includegraphics* [width=8.0cm]{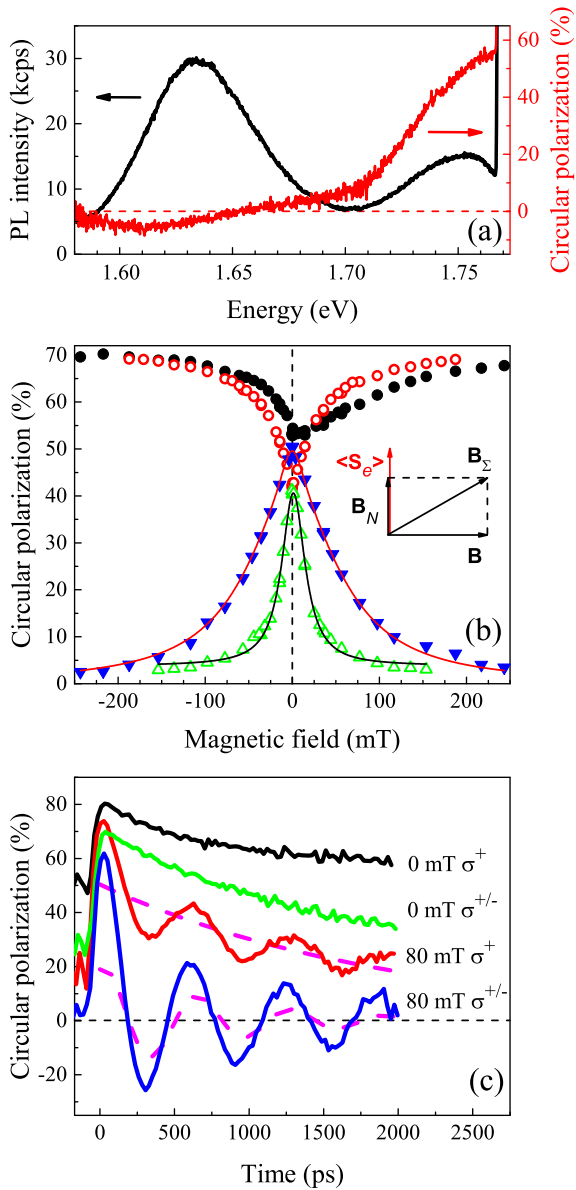}
\caption{(a) Spectra of the PL intensity (black line) and circular polarization (red line). (b) Dependence of the PL circular polarization degree on magnetic field in Voigt (triangles) and Faraday (circles) geometry. The open/filled symbols denote the results obtained in absence/presence of DNP. The black/red line denotes the fitting by one/two Lorentz curves in case of absence/presence of DNP. The inset shows the orientation of the mean electron spin $\langle \mathbf{S}_e \rangle$ and the magnetic fields relative to each other. (c) Dynamics of the PL circular polarization. Solid lines represent measured dependencies in zero magnetic field and in a Voigt field of 80~mT. For each field value the curves with and without DNP are presented. Dashed lines show curves obtained by fitting of the experimental data, the sum of which describes well the 80~mT curve obtained in presence of DNP.}
\label{fig1}
\end{figure}

The excitation photon energy (1.77~eV) was smaller than the band gap of the In$_{0.48}$Ga$_{0.52}$P barrier (1.96~eV). As a result, the carriers were generated in  excited QDs states. The two PL spectral bands in Fig.~\ref{fig1}(a) correspond to the two characteristic QD sizes~\cite{Samp}. The relatively small dots give rise to the band with maximum at 1.75~eV, while the large dots give rise to the band at 1.63~eV. In this work, we focus on the small QDs. All measurements [except the spectrum in Fig.~\ref{fig1}(a)] were performed at a detection photon energy of 1.75~eV. The PL of the small QDs has a significant degree of circular polarization ($40-50$\%) in zero magnetic field [Fig.~\ref{fig1}(a,b)], which is a signature for X$^{+}$ or X$^{-}$ trion PL~\cite{BasicInP}. We note that for these quantum dots optical orientation of the neutral exciton PL is almost absent~\cite{FastRel}.

Let us compare the magnetic field dependencies of the steady-state PL polarization in presence and in absence of DNP [Fig.~\ref{fig1}(b)]. For the latter case, we recall that the polarization of the exciting laser light was modulated in order to exclude DNP. In presence of DNP the half width at half maximum (HWHM) of the Hanle curve is three times larger (50~vs.~16~mT) compared to the case of absence, the polarization in zero magnetic field is also larger (48~vs.~41\%), and the curve of polarization recovery in Faraday field is asymmetric with respect to an inversion of the field sign. Here, the Hanle curves were fitted by single Lorenz curves for HWHM determination. The scenario observed in Voigt field contradicts the typical situation when the Overhauser field narrows the Hanle curve enhancing the effect of the external field and thus increasing the depolarization rate~\cite{OO5}. To describe the observed effects, it is straightforward to assume that the DNP takes place also in zero external magnetic field. As a result, the effective nuclear field stabilizes the electron spin. In addition, it can be assumed that even in presence of the Voigt field, the Overhauser field is oriented along the growth axis of the heterostructure. In this case, the Voigt field acting on the electron spins has to overcome the nuclear field in order to depolarize the PL. The asymmetry of the dependence in Faraday field arises from the fact that for one sign of the external field the nuclear field enhances it, while for the other sign the nuclear field reduces it. The reason for stabilization of the nuclear spins along the growth axis of the QDs is the nuclear quadrupole interaction caused by the lattice deformation, as will be discussed in Section~\ref{subsec:QI}.

Let us compare the time dependences of the PL circular polarization in presence and absence of DNP [Fig.~\ref{fig1}(c)]. At zero magnetic field, in presence of DNP the PL polarization is larger than in DNP absence (maximum is 80~vs.~70\%). As mentioned above, even in zero field the electron spins are stabilized by the nuclear field. When the external magnetic field is switched on in Voigt geometry, pronounced oscillations in the dynamics of the PL polarization are observed. The oscillations correspond to the Larmor precession of the electron spin contributing to the X$^{+}$ trion. In presence of dynamic polarization of the nuclear spins, the PL polarization oscillates not around zero value, but around some finite polarization contribution which monotonically decays with time. In the polarization dynamics we therefore distinguish between the ``monotonically decaying'' and ``oscillating'' (around zero polarization) contributions, which together form the experimentally measured curve [dashed lines in  Fig.~\ref{fig1}(c)]. The existence of the ``monotonically decaying'' contribution to the PL polarization can be interpreted by assuming that the Overhauser field is pinned along the QD growth axis. In this case, the electron spin is affected by the total magnetic field with oblique orientation ($\mathbf{B}_{\Sigma}$), given by the sum of the nuclear ($\mathbf{B}_N$) and the external ($\mathbf{B}$) fields, see inset in Fig.~\ref{fig1}(b). Thus, there is a component of the mean electron spin normal to $\mathbf{B}_{\Sigma}$ (and precessing in it) and a spin component parallel to $\mathbf{B}_{\Sigma}$ (no precession occurs). As a result, there are ``oscillating'' and ``monotonically decaying'' contributions to the PL polarization.

\begin{figure}[!h]
\centering
\includegraphics* [width=8.0cm]{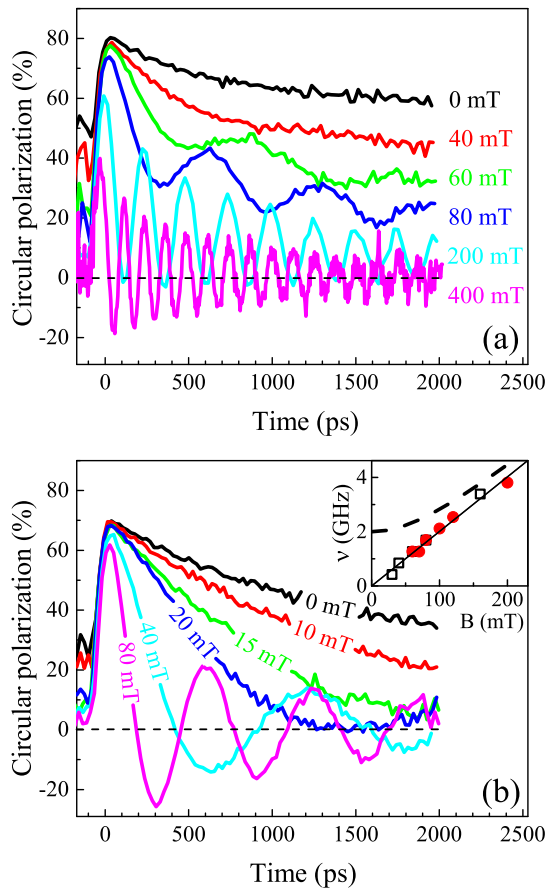}
\caption{Dynamics of the PL circular polarization in presence (a) and absence (b) of DNP in different Voigt magnetic fields. Inset in panel (b): dependence of the polarization oscillation frequency on magnetic field strength in presence (filled circles) and in absence (empty squares) of DNP. The calculated dependence is shown by the dashed line. }
\label{fig2}
\end{figure}

Figure~\ref{fig2} shows a series of PL polarization time dependencies in different magnetic fields measured in presence [Fig.~\ref{fig2}(a)] and in absence [Fig.~\ref{fig2}(b)] of DNP. Let us consider the results shown in Fig.~\ref{fig2}(a). We assume that the electron spin in the X$^{+}$ trion is affected by the nuclear Overhauser field pinned along the QD growth axis. The angle between the mean electron spin ($\langle \mathbf{S}_e \rangle$) and the total field ($\mathbf{B}_{\Sigma}$) increases with increasing external field ($\mathbf{B}$). As a result, the value of the spin projection on the $\mathbf{B}_{\Sigma}$ direction decreases, while the value of the spin projection on the axis normal to $\mathbf{B}_{\Sigma}$ increases. Thus, the amplitude of the ``monotonically decaying'' polarization contribution decreases with external magnetic field, as observed in Fig.~\ref{fig2}(a). The strength of the field at which the amplitude of the ``monotonically decaying'' contribution equals the amplitude of the ``oscillating'' contribution is approximately 100 mT. This value determines the strength of the effective field of the dynamically polarized nuclear spins ($\mathbf{B}_N$). In absence of DNP [Fig.~\ref{fig2}(b)] no noticeable ``monotonically decaying'' contribution is observed. Indeed, the electron spin projection on the direction of the external field is close to zero.

To determine the frequency of the electron spin Larmor precession, the signals shown in Fig.~\ref{fig2}(a,b) were Fourier-transformed. The oscillation frequency ($\nu$) increases linearly with external magnetic field both in presence and in absence of DNP [inset in Fig.~\ref{fig2}(b)]. The electron $g$-factor value, $\lvert g_e \rvert=1.43$, is determined from the slope of this linear dependence using $h\nu={\mu_B}\lvert g_e \lvert B$. It agrees with the value obtained in~\cite{X0}, while it differs from the value of 1.6 obtained in other studies~\cite{G2, G3}. However, if the nuclear field is pinned along the growth axis, the oscillation frequency will be proportional to  $\sqrt{B^2+B_N^2}$, see the dashed line in the inset of Fig.~\ref{fig2}(b). The model predicts a significant deviation from the straight line in fields smaller or equal to the nuclear field, ${B}_N=100$~mT, which we do not observe, see inset of Fig.~\ref{fig2}(b). Thus, in the range of small fields there is a contradiction between model and experiment. On the one hand, the magnetic field dependencies of the PL circular polarization indicate the presence of the nuclear field. On the other hand, the Larmor precession of the electron spin occurs as if the nuclear field is absent.

In order to resolve this contradiction, we propose that the PL consists of two independent contributions. The ``oscillating'' contribution corresponds to the X$^{+}$ trion, and the ``monotonically decaying'' contribution corresponds to the X$^{-}$ trion. Therefore, in the ensemble of nominally negatively charged QDs there is a subensemble of positively charged dots generated by photo-doping. A quantum well recharging under excitation below the barriers was reported for GaAs/(Al,Ga)As semiconductors~\cite{Volk}. In addition, to describe the experimental results it is necessary to assume that in presence of resident electrons (leading to X$^{-}$ trions) dynamic polarization of the nuclear spins takes place, while in absence of resident electrons (leading to X$^{+}$ trions) the DNP can be neglected. This point can be explained by the different lifetimes of resident and photoexcited carriers. In the latter case, there is most likely not sufficient time for significant interaction between the photoexcited and subsequently radiatively decaying electrons and the nuclei. Moreover, it is still assumed that the Overhauser field of the dynamically polarized nuclei (the X$^{-}$ case) is pinned along the QDs growth axis. 

Using these hypotheses we can consistently describe all experimental results. Let us consider now the oscillation frequency dependence on magnetic field [inset in  Fig.~\ref{fig2}(b)]. As noted in the introduction, the dynamics of optical orientation of X$^{-}$ trion PL does not exhibit oscillations in Voigt magnetic field (see also Sec.~\ref{subsec:Num}). Thus, the dependence is completely determined by the Larmor precession of the electron spin in the X$^{+}$ trion. In the case of the X$^{+}$ trion, absence of the DNP is proposed, so that the observed linear dependence of the frequency on magnetic field as well as the coincidence of the results in absence and presence of excitation modulation is expected. We recall that modulation of the excitation polarization at a sufficiently high frequency prevents DNP.

Let us consider the time dependencies of the PL polarization in presence of DNP [Fig.~\ref{fig2}(a)]. The polarization of the X$^{-}$ trion monotonously vanishes in Voigt magnetic field without observing oscillations. If one subtracts this contribution from the experimental signal for the X$^{+}$ trion, a polarization oscillating around zero value will remain [the dashed line in Fig.~\ref{fig1}(c)]. However, in absence of DNP [Fig.~\ref{fig2}(b)] the ``monotonically decaying'' contribution (related to the X$^{-}$ trion) is vanishing already in smaller fields (about 16~mT). In this case, the electron spin is influenced by the Voigt external magnetic field only, while in presence of DNP [Fig.~\ref{fig2}(a)] the Voigt external field has to be larger than the nuclear field, which direction coincides with the QDs growth axis, in order to depolarize the X$^{-}$ trion PL.

For the steady-state magnetic field dependencies of the polarization [Fig.~\ref{fig1}(b)] the arguments proposed earlier regarding the nuclear field that is pinned along the growth axis remain valid. At the same time, we assume that the Hanle curve, measured in absence of DNP, is the sum of two Lorentz curves, reflecting the contributions of the X$^{+}$ and X$^{-}$ trions, with approximately equal HWHM (16 mT) [Fig.~\ref{fig1}(b), the black line]. In presence of DNP, the Hanle curve can be fitted by the sum of two Lorentz curves with HWHMs of 16~mT and 70~mT [Fig.~\ref{fig1}(b), the red line]. Thus, the half width of the X$^{+}$ trion depolarization curve remains the same, while the X$^{-}$ trion depolarization curve is noticeably broadened due to the presence of the nuclear field. The strength of the nuclear field of 70~mT can be estimated from the HWHM of the Lorentz curve corresponding to the X$^{-}$ trion PL depolarization. 

\section{Discussion}
\label{sec:Disc}

\subsection {Nuclear quadrupole interaction}
\label{subsec:QI}

The pinning of the nuclear spins along the growth axis of the QDs is caused by the quadrupole interaction of the indium nuclei (95.5\% $^{115}$In and 4.5\% $^{113}$In, both with spin 9/2~\cite{NucSpin}). For the nuclei of phosphorus (with spin 1/2) the quadrupole interaction is absent. The nuclear quadrupole interaction plays a significant role in external magnetic field when the Zeeman splitting of nuclear spins is smaller than the quadrupole splitting. Estimates show that for uniaxial strain of 2\% directed along the QD growth axis the quadrupole splitting dominates over the Zeeman splitting up to 100~mT external magnetic field~\cite{Quadr}. In this field range the quadrupole effects are observed in our case.  Fields below 100~mT and oriented perpendicular to the quadrupole axis do not split doubly degenerate states which are associated with a fixed modulus of the nuclear spin projection (greater than 1/2) on the QD growth axis [Fig.~\ref{fig3}(a)], the degeneracy of these states is removed by the quadrupole interaction. As a result, the nuclear dipole-dipole interaction does not destroy the orientation of the dynamically polarized nuclear spins even in absence of an external magnetic field. It should be noted that the effect of nuclear quadrupole interaction on the system of electron and nuclear spins should manifest itself in different types of self-organized quantum dots with nuclear spins greater than 1/2. 

\subsection{Quantitative description of the PL polarization dynamics in absence of DNP}
\label{subsec:Num}

\begin{figure*}
  \centering
  \includegraphics[width=\textwidth]{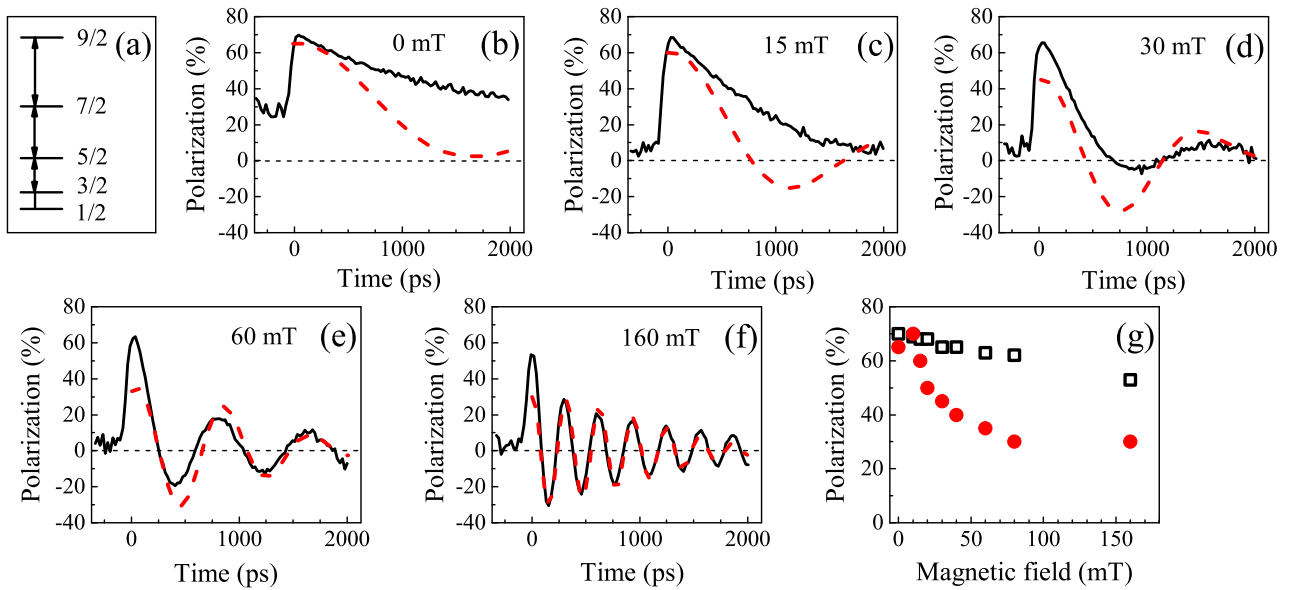}
  \caption{(a) Energy states corresponding to the modulus of the indium nuclei spin projection. (b - f) Time dependencies of PL circular polarization in various Voigt magnetic fields. The experiment is shown by solid lines, the modeling is shown by dashed lines. (g) Dependence of the experimental data (empty squares) and the calculated values (filled circles) of the polarization maximum on external magnetic field.}
  \label{fig3}
\end{figure*}

As can be seen in Fig.~\ref{fig1}(b), the widths of the Hanle curve and the polarization restoration curve in Faraday magnetic field are equal in absence of DNP. This is a characteristic feature of electron spin relaxation in nuclear field fluctuations with long correlation time. In this case there is a well-proven theory for describing the time dependencies of polarization~\cite{Merc}. The theory describes the QD ensemble-averaged electron spin relaxation in ``frozen'' nuclear field fluctuations subject to an external magnetic field. No recombination or other spin relaxation mechanisms are taken into account. The results of the polarization dynamics fitted by the corresponding functions are presented in Fig.~\ref{fig3}(b-f). The parameters of the model are the characteristic value of nuclear field fluctuations ($\Delta_{b}$), the modulus of the electron $g$-factor ($\lvert g_e \rvert$) and the characteristic value of the electron spin dephasing time ($T_{\Delta}$). Moreover, it is sufficient to know any two of these parameters to determine the third one. As noted above, $\lvert g_e \rvert=1.43$ was obtained from the experiment. $\Delta_b$ was chosen as fitting parameter. In addition, the functions were multiplied by the factor $A$ normalizing the amplitude. The best match is obtained for $\Delta_{b}=12$~mT, which is close to the HWHM of the Hanle curve (16~mT). It should be noted that in the limit of long correlation time of the field fluctuations the HWHM of the Hanle curve is determined by the strength of the nuclear field fluctuations. Based on the known dimensions of the QD ($100 \times 5$ nm$^{2}$) and the electron $g$-factor, one obtains a theoretical estimate of the value of the nuclear field fluctuations $\Delta_{b}$ of 20~mT~\cite{Merc}, which agrees well with the value experimentally obtained in this work (16~mT).

Let us turn to the data in zero magnetic field [Fig.~\ref{fig3}(b)]. After 100~ps from the moment of excitation the circular polarization degree reaches its maximum, which is equal to 70\% (but not 100\%). This indicates the presence of a fast spin relaxation mechanism (during the thermalization of carriers to the trion ground state). The polarization magnitude then decays in a few nanoseconds by a factor of 2.8 from 70 to 25\% where it remains constant until the next laser pulse comes in (in the figure one can see the constant level at ``negative delays'', the pulse repetition period is 13 ns). This situation is characteristic of spin relaxation in ``frozen'' fields. However, in magnetic fields smaller than 60~mT there is a significant mismatch between theory and experiment. Good agreement is achieved only in relatively large magnetic fields exceeding 60~mT [Fig.~\ref{fig3}(e,f)]. 

The presence of two contributions to the PL (X$^{+}$ and X$^{-}$) could be the reason for the mismatch between theory and experiment in small fields. In the singlet state of the X$^{+}$ trion the spins of the two holes compensate each other, so that the electron spin in the QD determines the trion spin during its lifetime. In this case, the model~\cite{Merc} can be used. In case of the X$^{-}$ trion, the nuclear field acts on the resident electron spin only until the electron pair singlet state with zero total spin is formed by photoexcitation. Then the spin dynamics is determined by the spin of the heavy hole in the trion. Thus, the case of X$^{-}$ trion is beyond the scope of the theory, and, as a result, the description of the experimental data in low fields faces a problem. In large fields (exceeding 40~mT) the ``monotonically decaying'' PL contribution of the X$^{-}$ trion vanishes, making the theory application possible. 

Let us consider the influence of a magnetic field in Voigt geometry on the spin dynamics of the X$^{-}$ trion. The transverse $g$-factor of the heavy hole is small ($g_{hh}^{\bot}~\ll~ 1$)~\cite{Ghole}. As a result, a Voigt magnetic field of about 100~mT does not affect the hole spin dynamics. In other words, the precession period of the hole spin and consequently of the polarization is longer than the hole lifetime. We recall that the orientation of the hole spin determines the PL polarization after  formation of the trion singlet state. Prior to that, the electron spins precess in magnetic field, and their orientation at the moment of singlet state formation affects the subsequent hole spin orientation. Now if there is a dispersion (over the QD ensemble) of the time of electron energy relaxation into the singlet state, the precession of the electron spins in different QDs will end at different times, so they will be differently oriented at the moment of thermalization. As a result, the average hole spin at the moment of singlet state formation will decrease with increasing field. The latter circumstance leads to the decrease (from 70 to 50\%) of the polarization maximum (obtained from the dynamic curves), which is reflected in the decrease (from 70 to 30\%) of the fit parameter $A$ [Fig.~\ref{fig3}(g)] that is introduced for normalization of the fit functions amplitude. 

In the field range $60-320$~mT the polarization dynamics are determined by the electron spin in the X$^{+}$ trion. As a result, good agreement between theory and experiment is achieved. In this case, the depolarization time ($T_{\Delta}$) does not depend on magnetic field, indicating the absence of a noticeable dispersion of the electron $g$-factor. 

\section{Conclusion}
\label{sec:conclusions}

We have shown that in presence of nuclear quadrupole interaction in InP/(In,Ga)P QDs the dynamically polarized spins of the nuclei and the spins of the electrons are stabilized when there are resident electrons in the QDs. Moreover, the stabilization takes place even in zero external magnetic field. In addition, the quadrupole interaction leads to pinning of the Overhauser field along the quadrupole axis, close to the QDs growth axis. The latter circumstance, as well as the presence of two contributions to the PL, one of which corresponds to the PL of the X$^{-}$ trion, and the other one to the X$^{+}$ trion, significantly affects the dynamics of circular polarization in Voigt magnetic field.

We have also shown that relaxation of the electron spins in the ``frozen'' fluctuations of the nuclear field takes place. The experimentally measured time dependencies of the PL polarization were modeled using the theory presented in Ref.~\cite{Merc}. In the range of relatively large magnetic fields ($60-320$~mT), good agreement between theory and experiment was achieved. The mismatch between theory and experiment in the field range up to 60~mT may be caused by the presence of the X$^{-}$ contribution to the PL polarization, which is outside of the scope of the model.

\section*{Acknowledgements}

The authors are grateful to V.~L.~Korenev and K.~V.~Kavokin for fruitful discussions and to M.~Salewski for his help with the experiment. S.~V.~Nekrasov  and Yu.~G.~Kusrayev acknowledge the support by the Russian Science Foundation (Grant No.~18-12-00352). I.~A.~Akimov, Yu.~G.~Kusrayev, D.~R.~Yakovlev and M.~Bayer acknowledge the support by the Deutsche Forschungsgemeinschaft via the project No.~409810106 and in the frame of the International Collaborative Research Center TRR 160 (Project B4) and by
the Russian Foundation for Basic Research (Grant No. 19-52-12066).


\begin{thebibliography}{}

\bibitem{Dyakonov}  X.~Marie, B.~Urbaszek, O.~Krebs, and T.~Amand, in \textit{Spin Physics in Semiconductors}, edited by  M.~I.~Dyakonov (Springer-Verlag, Berlin, 2008), Chap. 4.

\bibitem{Bracker} A.~S.~Bracker, D.~Gammon, and V.~L.~Korenev, Fine structure and optical pumping of spins in individual semiconductor quantum dots, Semicond. Sci. Technol. \textbf{23}, 114004 (2008).

\bibitem{OO3}  G.~E.~Pikus and A.~N.~Titkov, in \textit{Optical Orientation}, edited by  F.~Meier and B.~Zakharchenya (North-Holland, Amsterdam, 1984), Chap. 3.

\bibitem{Khaet} A.~Khaetskii and Yu.~V.~Nazarov, Spin relaxation in semiconductor quantum dots, Phys. Rev. B \textbf{61}, 12639 (2000).

\bibitem{Quadr} R.~I.~Dzhioev and V.~L.~Korenev, Stabilization of the Electron-Nuclear Spin Orientation in Quantum Dots by the Nuclear Quadrupole Interaction, Phys. Rev. Lett. \textbf{99}, 037401 (2007).

\bibitem{Slichter}  C.~P.~Slichter, \textit{Principles of Magnetic Resonance} (Springer-Verlag,  Berlin, 1990).

\bibitem{OO5}  V.~G.~Fleisher and I.~A.~Merculov, in \textit{Optical Orientation}, edited by  F.~Meier and B.~Zakharchenya (North-Holland, Amsterdam, 1984), Chap. 5.

\bibitem{Kurt} A.~Kurtenbach, K.~Eberl, and T.~Shitara, Nanoscale InP islands embedded in InGaP, Appl. Phys. Lett. \textbf{66}, 361 (1995).

\bibitem{Cheh} E.~A.~Chekhovich,  M.~N.~Makhonin, J.~Skiba-Szymanska, A.~B.~Krysa, V.~D.~Kulakovskii, M.~S.~Skolnick, and A.~I.~Tartakovskii, Dynamics of optically induced nuclear spin polarization in individual ${\rm InP/Ga_xIn_{1-x}P}$ quantum dots, Phys. Rev. B \textbf{81}, 245308 (2010).

\bibitem{XPlus} L.~Lombez, P.-F.~Braun, X.~Marie,  P.~Renucci, B.~Urbaszek, T.~Amand, O.~Krebs, and P.~Voisin, Electron spin quantum beats in positively charged quantum dots: Nuclear field effects, Phys. Rev. B \textbf{75}, 195314 (2007).

\bibitem{X0} I.~A.~Yugova, I.~Ya.~Gerlovin, V.~G.~Davydov, I.~V.~Ignatiev, I.~E.~Kozin, H.~W.~Ren, M.~Sugisaki, S.~Sugou, and Y.~Masumoto, Fine structure and spin quantum beats in InP quantum dots in a magnetic field, Phys. Rev. B \textbf{66}, 235312 (2002).

\bibitem{My} S.~V.~Nekrasov, Yu.~G.~Kusrayev, I.~A.~Akimov, V.~L.~Korenev, L.~Langer, and M.~Salewski, Negative circular polarization dynamics in InP/InGaP quantum dots, J. Phys.: Conf. Ser. \textbf{741}, 012189 (2016).

\bibitem{AEI}  E.~L.~Ivchenko and G.~E.~Pikus, \textit{Superlattices and Other Heterostructures} (Springer-Verlag,  Berlin, 1997).

\bibitem{FastRel} M.~Paillard, X.~Marie, P.~Renussi, T.~Amand, A.~Jbeli, and J.~M.~Gerard, Spin Relaxation Quenching in Semiconductor Quantum Dots, Phys. Rev. Lett. \textbf{86}, 1634 (2001).

\bibitem{ZeroOsc1} A.~P.~Heberle, J.~J.~Baumberg, and K.~Kohler, Ultrafast Coherent Control and Destruction of Excitons in Quantum Wells, Phys. Rev. Lett. \textbf{75}, 2598 (1995).

\bibitem{ZeroOsc2} T.~Amand, X.~Marie, P.~Le~Jeune, M.~Brousseau, D.~Robart, J.~Barrau, and R.~Planel, Spin Quantum Beats of 2D Excitons, Phys. Rev. Lett. \textbf{78}, 1355 (1997).

\bibitem{Merc} I.~A.~Merkulov, Al.~L.~Efros, and M.~Rosen, Electron spin relaxation by nuclei in semiconductor quantum dots, Phys. Rev. B \textbf{65}, 205309 (2002).

\bibitem{Samp} J.~Kapaldo, S.~Rouvimov, J.~L.~Merz, S.~Oktyabrsky, S.~A.~Blundell, N.~Bert, P.~Brunkov, N.~A.~Kalyuzhnyy, S.~A.~Mintairov, S.~Nekrasov, R.~Saly, A.~S.~Vlasov, and A.~M.~Mintairov, Ga-In intermixing, intrinsic doping, and Wigner localization in the emission spectra of self-organized InP/GaInP quantum dots, J. Phys. D: Appl. Phys. \textbf{49}, 475301 (2016).

\bibitem{BasicInP} A.~S.~Bracker, E.~A.~Stinaff, D.~Gammon, M.~E.~Ware, J.~G.~Tischler, A.~Shabaev, Al.~L.~Efros, D.~Park, D.~Gershoni, V.~L.~Korenev, and I.~A.~Merkulov, Optical Pumping of the Electronic and Nuclear Spin of Single Charge-Tunable Quantum Dots, Phys. Rev. Lett. \textbf{94}, 047402 (2005).

\bibitem{G2} A.~A.~Sirenko, T.~Ruf, A.~K.~Kurtenback, and K.~Eberl, in \textit{23rd International Conference on the Physics of Semiconductors} (World Scientific,  Berlin, 1996), Vol. 2, p. 1385.

\bibitem{G3} M.~Syperek, D.~R.~Yakovlev, I.~A.~Yugova, J.~Misiewicz, M.~Jetter, M.~Schulz, P.~Michler, and M.~Bayer, Electron and hole spins in InP/(Ga,In)P self-assembled quantum dots, Phys. Rev. B \textbf{86}, 125320 (2012).

\bibitem{Volk} O.~V.~Volkov, I.~V.~Kukushkin, D.~V.~Kulakovskii, K.~von~Klitzing, and K.~Eberl, Bistable Charge States in a Photoexcited Quasi-Two-Dimensional Electron-Hole System, JETP Lett. \textbf{71}, 322 (2000).

\bibitem{NucSpin}  A.~ L\"oshe, \textit{Kerninduktion} (veb Deutscher Verlag der Wissenschaften,  Berlin, 1957).

\bibitem{Ghole} J.~G.~Tischler, A.~S.~Bracker, D.~Gammon, and D.~Park, Fine structure of trions and excitons in single GaAs quantum dots, Phys. Rev. B \textbf{66}, 081310(R) (2002).

\end{thebibliography}
\end{document}